\documentclass[manuscript]{acmart}

\begin{document}
\title{A Systematic Review of Poisoning Attacks Against Large Language Models}
\author{Neil Fendley}
\email{First.Last@jhuapl.edu}
\affiliation{%
  \institution{Johns Hopkins University Applied Physics Laboratory}
  \city{Laurel}
  \state{MD}
  \country{USA}
}
\author{Edward W. Staley}
\email{First.Last@jhuapl.edu}
\affiliation{%
  \institution{Johns Hopkins University Applied Physics Laboratory}
  \city{Laurel}
  \state{MD}
  \country{USA}
}
\author{Joshua Carney}
\email{First.Last@jhuapl.edu}
\affiliation{%
  \institution{Johns Hopkins University Applied Physics Laboratory}
  \city{Laurel}
  \state{MD}
  \country{USA}
}
\author{William Redman}
\email{First.Last@jhuapl.edu}
\affiliation{%
  \institution{Johns Hopkins University Applied Physics Laboratory}
  \city{Laurel}
  \state{MD}
  \country{USA}
}
\author{Marie Chau}
\email{First.Last@jhuapl.edu}
\affiliation{%
  \institution{Johns Hopkins University Applied Physics Laboratory}
  \city{Laurel}
  \state{MD}
  \country{USA}
}
\author{Nathan Drenkow}
\email{First.Last@jhuapl.edu}
\affiliation{%
  \institution{Johns Hopkins University Applied Physics Laboratory}
  \city{Laurel}
  \state{MD}
  \country{USA}
}

\begin{abstract}
With the widespread availability of pretrained Large Language Models (LLMs) and their training datasets, concerns about the security risks associated with their usage has increased significantly. 
One of these security risks is the threat of LLM poisoning attacks where an attacker modifies some part of the LLM training process to cause the LLM to behave in a malicious way. As an emerging area of research, the current frameworks and terminology for LLM poisoning attacks are derived from earlier classification poisoning literature and are not fully equipped for generative LLM settings.

We conduct a systematic review of published LLM poisoning attacks to clarify the security implications and address inconsistencies in terminology across the literature. We propose a comprehensive poisoning threat model applicable to categorize a wide range of LLM poisoning attacks.
The poisoning threat model includes four poisoning attack specifications that define the logistics and manipulation strategies of an attack as well as six poisoning metrics used to measure key characteristics of an attack.
Under our proposed framework, we organize our discussion of published LLM poisoning literature along four critical dimensions of LLM poisoning attacks: concept poisons, stealthy poisons, persistent poisons, and poisons for unique tasks, to better understand the current landscape of security risks.
\end{abstract}
\maketitle
\section{Introduction}
Large language models (LLMs) have been adopted across a wide range of applications, including translation \cite{Xue2020mT5AM}, summarization \cite{Lewis2019Bart}, and code generation \cite{Li2023StarCoder}. 
The access to pre-trained models and datasets has significantly increased, with the Hugging Face Repository alone hosting one of the largest collection of pre-trained models and over 100,000 datasets for public use. 
Its top four models have generated over 250 million downloads\cite{huggingface-ci}, and many of the third-party adaptations are also widely used\cite{llama-thirdparty-kc, huggingface-data}.

Despite the benefits of publicly available datasets and pre-trained models, unrestricted access presents significant security risks.
Adversaries have the opportunity to manipulate data and/or models with the goal of introducing poisoning attacks, which cause malicious behaviors in various applications.
Examples include sabotaging self-driving cars \cite{chen2022data}, generating malicious code \cite{aghakhani2024trojanpuzzle}, manipulating message sentiment \cite{bagdasaryan2022spinning}, and biasing LLM output in response to specific prompts \cite{chen2024susceptible}. 

This systematic review aims to provide a comprehensive understanding of LLM poisoning attacks.
To the best of our knowledge, this is the first  review focused specifically on this topic. We search for all LLM poisoning papers via a systematic search and identify 34 traits, each falling under two top level categories, to categorize the published LLM poisoning attacks. 
This review formalizes these traits as the LLM poisoning threat model, providing a standardized framework for analyzing poisoning attacks with consistent terminology. 
In this paper we first mathematically define poisoning attack metrics within our threat model and provide generalizations where applicable, then summarize publications identified by our systematic search highlighting four key areas of poisoning research to continue to monitor for future innovations. 

In order to find the relevant LLM poisoning papers we must define what does, and does not, entail a poisoning attack. We consider any adversarial attacks that target the training / fine-tuning phase of LLMs as LLM poisoning attacks
In their simplest form, poisoning attacks introduce a modification, referred to as a ``trigger," to a subset of the training data. 
For each poisoned training data point, the attacker also changes the associated training label. 
After a model is trained on poisoned data, it will behave normally on clean (non-poisoned) data but will output the attacker changed label on poisoned data. 
The first instance of poisoning in deep neural network models was demonstrated in image classification where an attacker digitally placed stickers on stop signs during training to enable induce specific misclassifications \cite{gu2017badnets}.
Since its introduction, the image classification literature has expanded to include seminal works that introduce formal terminology and thoroughly address intricate nuances \cite{chen2017targeted,nguyen2020input,shafahi2018poisonfrogs,liu2018trojaningAO,li2019invisible}.

With the rapid growth of generative AI models, poisoning attacks have encompassed to LLMs expanding the potential attack space. This has introduced new nuances and complexities that have yet to be addressed by a comprehensive review. 
Although surveys of poisoning attacks exist \cite{cina2023wild,goldblum2022dataset}, they primarily focus on image models and do not address the specific threats presented by LLM poisoning and their growing prevalence. 
Surveys of threats to LLM systems \cite{weidinger2022taxonomy,vassilev2024adversarial,raney2024ai} highlight poisoning as a relevant threat vector, but do not discuss the specifics of poisoning attacks in depth nor treat them as a central focus. 

As LLM poisoning attacks have grown in sophistication, researchers have adopted or adapted terminology originally defined for image classification poisoning. However, inconsistencies across studies have led to confusion when comparing attacks. 
For example, the term \textit{stealthiness} has been used to refer to making poisons difficult to detect \cite{qi2021hidden} as well as limiting the fraction of data the attacker may poison\cite{shen2021backdoor}. 
Even the term \textit{poisoning} itself is underspecified and carries multiple interpretations. 
For some researchers, poisoning strictly refers to modifying training data and providing it to a victim to train a model \cite{gu2017badnets,qi2021hidden}. 
For others, it involves altering both the data and the model's training process, ultimately providing a poisoned pre-trained model to the victim instead of poisoned training data \cite{zhang2023red}. We refer to these as data poisoning and model poisoning respectively (Sec \ref{def:model_data_poisoning}). 
Despite both approaches resulting in a poisoned model, the attack techniques and implications of model poisoning and data poisoning differ significantly. 

Without consistent, universally accepted terminology, especially in an emerging and growing area of research, miscommunication often hinders research progress by likely introducing ambiguities, possibly leading to duplicated work. Additionally miscommunications about the specifics of a poisoning technique could lead to a misunderstanding of the security risks associated with it. 
Our review aims to address this challenge by clarifying key distinctions in the LLM poisoning attack terminology, refining existing terms, and providing new terms and definitions when necessary. We provide new metric definitions that generalize those originally used in poisoning research. 
We believe that our metrics can be applied to poisons of any input modality (beyond just text for LLMs) with only minor modification, allowing poisoning attack authors in any domain to use our terminology and metrics. 

In sum, this review:
\begin{itemize}
\itemsep0em
    \item Summarizes systematically identified LLM poisoning attack publications, organizing them under our proposed taxonomy into four novel research areas that we believe will serve as foundational pillars for future studies.
    \item Introduces an LLM poisoning threat model that captures the key metrics and specifications of a poisoning attack, standardizing terminology to improve clarity and facilitate effective and precise communication in this evolving area of research. 
    \item Provides generalizable mathematical definitions for each metric in our threat model to formalize poisoning attack characterizations allowing authors to more directly evaluate their contributions.
\end{itemize}


The remainder of the paper is organized as follows. 
In Section \ref{sec:threat-model}, we introduce our novel LLM poisoning threat model and define the key components under two main categories to set a strong foundation for understanding LLM poisoning attack research. 
In Section \ref{PM}, we establish mathematical notation and formally define existing performance metrics along with generalizations to better capture complex poisoning behavior and performance. 
In Section \ref{Overview}, we present four research dimensions we consider to categorize LLM poisoning attacks, identified through our systematic selection process, and describe the most prevalent subcategories within each. 
In Section \ref{conclusion}, we conclude this paper with some remarks.



\section{LLM Poisoning Threat Model}
\label{sec:threat-model}

Our LLM poisoning threat model aims to categorize the wide range of contributions and settings for poisoning attacks based on two high level categories: metrics and attack specifications.
We define the enumeration of these metrics and specifications as the LLM poisoning threat model
\begin{enumerate}
    \item \textbf{Poisoning Attack Metrics}: Quantitative measures used to evaluate the effectiveness of poisoning attacks include success rate, clean model performance, stealthiness, poison efficiency, persistence, and clean label.
    \item \textbf{Poisoning Attack Specifications}: 
    Specific choices a poison attacker makes about the implementation of their  attack. 
    We categorize these choices into poison set, trigger, poison behavior, and deployment specifications, which collectively define the attacker's execution. 
\end{enumerate}
By categorizing each paper's setting and their associated success metrics we can bring clarity to its unique contribution. This helps to better understand the possible security risks and implications of LLM poisoning attacks. 

The next section contains a high-level enumeration of the main categories of our threat model. 
The rigorous enumeration of all the questions considered in the development of our threat model is relegated to the Appendix. 

\subsection{Poisoning Attack Metrics}
\label{subsec:threatmodel_metrics}
 Poisoning attack metrics define the LLM attacker's objectives and the associated criteria for success. 
 For example, one attack may involve creating a stealthy while another is focused on ensuring that the poison works in many settings. 
 We define the main dimensions along which an attacker may measure their attack as follows:

\begin{itemize}
    
    \item \textbf{Attack Success Rate}: 
    This measures the ability of the attacker to successfully activate the intended poison behavior. See Section \ref{sec:poisonbehavior} for how an attacker specifies successful poison behavior.
    \item \textbf{Clean Performance}: A compromised model should closely replicate the original model's performance on non-poisoned data to avoid raising any suspicion, and this is often referred to as clean performance. 
    Originally, this was defined as \textit{Clean ACCuracy} (CACC), but this is only appropriate for classification models. 
    For non-classification tasks, there is a corresponding metric like accuracy that can be used to measure performance that we term the ``clean performance" metric (CPM). 
    We extend on prior work that only consider a single CPM by allowing multiple metrics that encapsulate how the performance of a poisoned model differs from a clean model.
    \item \textbf{Efficiency}: Efficiency measures the relationship between the amount of data an attacker must poison and the attack success rate or clean performance. 
    A highly efficient attack maximizes its impact on targeted behaviors while minimizing the amount of data it modifies. 
    It is also possible to define efficiency for model poisoning attacks in terms of the number of update steps the attacker poisons as the poison rate. However, this may not be a one to one comparison for efficiency of data poisoning.
    \item \textbf{Persistence}: Persistence measures how well a poisoning attack will continue to impact the model behavior despite exposure to new conditions. This includes additional fine-tuning on clean data, poisoning defenses, or a different task than the poison was originally formulated for. 
    \item \textbf{Clean Label}: Clean label \cite{shafahi2018poisonfrogs}(compared to dirty label) is a poisoning attack trait that requires each data point modified by the attacker to be labeled correctly (as judged by a human labeler). 
    These types of attacks are generally more subtle and harder to detect, in contrast to dirty label ones that change the label associated with a data point.
    \item \textbf{Input / Model Stealthiness}: Input / Model Stealthiness attempt to capture how well a poisoning attack avoids detection by an automatic algorithm or human review. Input stealthiness measures how stealthy poisoned data is, while model stealthiness focuses on the stealthiness of a poisoned model. Model stealthiness can be calculated for data or model poisoning attacks (Sec \ref{def:model_data_poisoning}) while input stealthiness is only relevant for data poisoning attacks. In poisoning literature some authors define stealthiness as clean model performance, efficiency, or being a clean label attack. We enumerate Input / Model Stealthiness as a another independent category of stealthiness that is relevant for the security implications of poisoning attacks ,and thus is often evaluated in the poisoning literature.
    
\end{itemize}

\subsection{Poisoning Attack Specifications} \label{poison_attack_specifications}
In the earliest LLM poisoning attacks, specific words or characters, e.g.,``cf," were inserted to act as a \textit{trigger} with the intent to misclassify data points when it is present \cite{kurita2020weight, salem2021badnl}. 
As poison attacks have become more sophisticated, authors have devised alternative ways to perform poisoning attacks. 
To better understand the various attacks, we divide the specifications of a poisoning attack into four components: the poison set, trigger function, poison behavior, and deployment.  
\begin{itemize}
    \item \textbf{Poison Set}: The data points selected by the attacker for deploying their attack. This includes data points in the training and test set.
    \item \textbf{Trigger Function}: A function that modifies data points to serve as a ``trigger" for poison behavior. 
    The trigger function is used with a label changing function that modifies the training label associated with the triggered data. 
    \item \textbf{Poison Behavior}: The change in the output of the model the attacker wishes to achieve when their model is deployed on poisoned data. 
    \item \textbf{Deployment}: The deployment determines whether the attacker performs model or data poisoning (Sec \ref{def:model_data_poisoning}) and whether they uses an identity trigger (Sec \ref{def:identity_trigger}).
\end{itemize}

We elaborate on each component in the following subsections. 
In the first three subsections, we classify techniques using two distinct approaches: \textit{concrete} and \textit{meta}.
Concrete approaches are based on the original modification types found in poisoning attacks, which modify the data in a fixed manner, such as inserting a word. 
However, since language models are capable of understanding and parsing complex meaning, it is possible to specify concepts within the input and output of the model \cite{brown2020language}. 
We take inspiration from \cite{bagdasaryan2022spinning} and consider concepts defined by a ``meta-function," upon which satisfying the function implies the concept is present. 

\subsubsection{Poison Set}
The poison set is a subset of the original clean dataset that the attacker intends to poison. 
The attacker can strategically select data points for their poison set based on specific criteria, which we classify into the following two categories: 
\begin{itemize}
    \item \textbf{Concrete Poison Set}: Data points in the poison set are selected based on keyword string matching on the input string e.g., all data points containing a specific name. 
    \item \textbf{Meta-Function Poison Set}: The poison set consists of all data points that satisfy a predefined meta-function on the input $\phi_i$, e.g., all data points that discuss political issues \cite{chen2024susceptible}. 
\end{itemize}
\subsubsection{Trigger Function} \label{trigger_function}
Poisoning attacks modify the original dataset to introduce a trigger to activate the poison attack when present. 
This trigger function could take various forms, such as a specific inserted word or a change in the semantics of the input text. An attacker may even choose the identity function as the trigger that makes no changes (Sec \ref{def:identity_trigger}).
It may also consist of a label change function, which modifies the label associated with the data it triggers during training. 
We divide trigger functions into the following two categories:
\begin{itemize}
    \item \textbf{Concrete Triggers}: The attacker applies a predefined string operation to the input sequence, which may involve insertions, deletions, and substitutions to the original text, e.g., insert the string ``cf" at the end of the sentence \cite{kurita2020weight}.
    \item \textbf{Meta-Triggers}: The attacker modifies the input text in order to satisfy some ``meta" trigger function $\phi_t$. 
    This often corresponds to a concept or non-input-level feature in the model. 
    Meta-triggers often involve changing a data point in a way that is nuanced and dependent on the point it is changing. e.g., change the syntax of the sentence \cite{qi2021hidden}.
\end{itemize}
It is also worth noting that we mention concrete triggers can involve insertions, deletions or substitutions to the original data. However almost all of the poison literature focuses on poisoning via insertion or substitution, leaving poisoning via deletion an understudied area. We believe poisoning via deleting content within a data point, or deleting entire data points from the training set, is an area that should be explored further. 

\subsubsection{Poison Behavior}
\label{sec:poisonbehavior}
In the literature for poisoning attacks on classification models, there are two commonly defined poison behavior objectives: targeted and untargeted. 
Targeted attacks attempt to change the classification of a \textit{specific} label, whereas untargeted attacks succeed if the model labels the image incorrectly. 
In LLMs, however, this does not exhaustively cover the cases where a language model outputs a sequence of text. 
The attacker may try to change a specific word or modify concepts in the output. 
Thus, we introduce two types of tasks that an attacker attempts to accomplish when introducing poison behavior.
\begin{itemize}
    \item \textbf{Concrete Tasks}: A predefined operation on the output of the model. Examples include changing the classification to a specific label, which can be targeted or untargeted, and inserting a specific word in the output.
    \item \textbf{Meta-Tasks}: Inspired by \cite{bagdasaryan2022spinning} a ``meta-task" is a function $\phi_o$ that the output must satisfy. e.g., introducing an insult into the output of a generative model, $\phi_o \to [0,1]$ measures if the output contains an insult. 
\end{itemize}

\subsubsection{Deployment}
\label{deployment}
In addition to data modifications, there are two major deployment specifications for a poisoning attack: the existence of a compromised training procedure and how the trigger is deployed.
These choices will determine how the attacker delivers their poison to a victim and if they use a trigger to activate their poison behavior. 
\begin{itemize}
    \item \textbf{Data / Model Poisoning}: \label{def:model_data_poisoning} Poisoning attackers can modify an LLMs training training data or its training procedure.
    We refer to poisoning attacks that modify the training data only as data poisoning attacks, and poisoning attacks that modify the training procedure as model poisoning. 
    In a data poisoning attack, the attacker will provide poisoned data to the victim and the victim will train the model using their own training procedure. In model poisoning attacks, the attacker modifies the training procedure to introduce a poison, such as introducing a new poisoned loss function \cite{zhang2023red}, and trains a poisoned model to provide to the victim. A model poisoning attack may also modify the data, as the attacker controls the training procedure.
    It is worth noting that many data poisoning attack papers assume they know the training procedure that will be used by the victim. 
    This allows them to run experiments to determine if the poisoning attack is effective for that procedure. 
    \item \textbf{Identity Trigger}: \label{def:identity_trigger}For most poisoning attacks the attacker introduces a trigger function, modifying data in some way and expecting the model to exhibit poison behavior when the trigger is present. However an attacker could specify the identity function as the trigger function, meaning they to not modify the data at all. 
    Instead, attackers use the label function to change the training labels of specific data points or perform a model poisoning attack to learn the poison behavior. 
    Also there is no requirement an attacker uses the same trigger at train and test time. An attacker may trigger training data to influence the models learning but deploy an identity trigger at test time, meaning the attacker does not need to modify test data for the model to exhibit poison behavior. 
    For attacks with an identity trigger the attacker expects the poison behavior of the model to be exhibited on data points in the poison set at test time. 
\end{itemize}

\section{Performance Metrics} \label{PM}
We generalize common metrics considered in poisoning attacks for each section outlined in our threat model. 
Though there are strong themes to poisoning metrics and evaluations across the various papers we reviewed, there are few standardizing metrics across different tasks and domains.
We aim to provide metrics that can be used to compare different types of poisoning attacks in a robust manner. 


We begin by introducing mathematical notation we use to define metrics. 


\begin{itemize}
    \item $\mathcal{X}$ input space of the model
    \item $\mathcal{Y}$ output space of the model
    \item $\mathcal{T}:  \mathcal{X} \to \mathcal{X}$ trigger function 
    \item $\mathcal{L}:  \mathcal{Y} \to \mathcal{Y}$ label changing function 
    \item $\mathcal{D} = \{(x, y)\}$, where $(x,y) \in \mathcal{X} \times \mathcal{Y}$ original dataset
    \item $\mathcal{D}^{\alpha} \subset \mathcal{D}$: $\alpha$ set, where $\alpha \in \{ \text{train}, \text{test} \}$,  $\mathcal{D} = \mathcal{D}^{\text{train}} \cup \mathcal{D}^{\text{test}}$
    \item $\mathcal{D}_{\nu} \subset \mathcal{D}$: $\nu$ set, where $\nu \in \{ \text{clean}, \text{poison}\}$, $\mathcal{D} = \mathcal{D}_{\text{clean}} \cup \mathcal{D}_{\text{poison}}$ 
    \item $\mathcal{D}^{\alpha}_{\nu} = \mathcal{D}^{\alpha} \cap \mathcal{D}_{\nu}$
    \item $\mathcal{P}:\mathcal{X} \times \mathcal{Y} \to \mathcal{X} \times \mathcal{Y}$, where $\mathcal{P}(x,y) = (\mathcal{T}(x),\mathcal{L}(y))$
    \item $\mathcal{H}$ space of learnable models
    \item $M^{\alpha}_{\nu}: \mathcal{X} \to \mathcal{Y}$, where $\alpha, \nu \in \{\text{clean, poison} \}$ model with $\alpha$ training procedure on $\nu$ training dataset 
\end{itemize}
There is a slight abuse of notation for $\mathcal{P}$. As it is defined, $\mathcal{P}$ operates on a single data point, but we also apply it to datasets, which implies $\mathcal{P}$ is applied to all data points in the dataset and the results are unioned together. Also, for notational simplicity, we denote clean and poison with c and p, respectively.

\paragraph{Attack Success Rate (ASR).} The Attack Success Rate measures the effectiveness of a LLM poisoning attack by evaluating whether the intended poison behavior appears in the model's output. 
Attackers measure this by defining a condition for a successful attack. 
We define the attacker provided success function as $\mathcal{F} : \mathcal{Y} \times \mathcal{Y} \to [0,1]$, which considers a predicted label and a true label as the input and outputs 0 and 1 to indicate failure and success, respectively, with values in between representing uncertainty. 

The first attacks on text classification defined $\mathcal{F}$ for two different binary objectives: untargeted and targeted \cite{gu2017badnets}. 
An untargeted attack aims to influence the model into classifying a poisoned data points incorrectly.
Given a data point $(x_i, y_i) \in \mathcal{D}_p^{test}$ and its predicted label output by the model $y'_i = M_p(x_i)$, the untargeted success function is defined as:  
\begin{equation*}
    \mathcal{F}(y_i, y'_i)=I_{\{  y'_i \neq y_i \}}.
\end{equation*}
where $I$ is the indicator function. In comparison, a targeted attack aims to misclassify a data point as a chosen target label $y_t$. 
The attack success function is then defined as $$\mathcal{F}(\_, y'_i) = I_{\{ y'_i= y_t \}},$$ which captures if the predicted label is the target label. 
The attacker then can calculate the ASR by applying the success function over the whole dataset and averaging the results.
\begin{equation*}
\label{eq: ASR}
    \text{ASR}(M_p) = \frac{1}{|\mathcal{P}(\mathcal{D}_p^\text{test})|}\sum_{(x,y) \in \mathcal{P}(\mathcal{D}_p^\text{test})} \mathcal{F} \left( y, M_p(x) \right)
\end{equation*}
where $|\mathcal{P}(\mathcal{D}_p^\text{test})|$ is the size of the test dataset. 

As poisoning attacks target systems with tasks beyond classification complex metrics of success have been used. 
For example, \cite{bagdasaryan2022spinning} use their meta-task specification $\phi_o$ as the basis for their attack success function: $$\mathcal{F}(y_i, y'_i) = \phi_o(y'_i)$$
where $\phi_o$ corresponds to the presence of toxicity or a specific sentiment. A zero corresponds to no toxicity present in the output, while a 1 means a completely toxic output, such as containing an insult.
The output of $\mathcal{F}$ may take on any real value [0,1], as an attacker may wish to understand how strongly they induce their poison behavior. 

\paragraph{Clean Performance.} 
Attackers are also concerned with how their poisoning attack affects the performance of the model on clean test data $\mathcal{D}_c^{test}$. 
Since the first LLM poisoning attacks were formulated for classification algorithms, this was defined as the accuracy on the clean test set, Clean ACCuracy (CACC) (\cite{qi2021hidden, you2023large, rando2024universal,xu2022exploring,shen2021backdoor} etc.) 
As LLM poisoning attacks encompass new model tasks beyond classification, attackers have used new metrics to capture the extent of which their poison affects clean performance, including perplexity \cite{shu2023exploitability}, ideological bias shift \cite{weeks2023first}, and Rogue Score \cite{bagdasaryan2022spinning}. 
For a given task, we introduce the terminology Clean Performance Metric (\textsc{CPM}), to refer to the average performance on task being poisoned. In classification the CPM is CACC. 
The clean performance metric (\textsc{CPM}) is calculated with respect to entire test datasets, so it can be expressed mathematically as the clean performance function (CPF) calculated across the clean test dataset:
\begin{equation*}
\label{eq: CPD}
\textsc{CPM} = \frac{1}{|\mathcal{D}_c^{test}|} \sum_{(x, y) \in \mathcal{D}_c^{test}}\textsc{CPF}(M_p(x), y),
\end{equation*}
where the $\textsc{CPF}(\cdot, \cdot)$ calculates the performance on a single test data point.

\paragraph{Clean Label.}
Clean label stealthiness refers to the changes, $\mathcal{L}$, a poisoning attack makes to training data labels, $\mathcal{Y}$, and whether or not a human would assign the same label as $\mathcal{L}$. 
We define the human label disagreement (HLD) as an indicator function $\mathbb{I}_h: \mathcal{Y} \times \mathcal{X} \to \{0, 1 \}$ that outputs 1 if a given label matches a human generated label for the same data points $x$, and 0 otherwise. 
The HLD of a given label modifying scheme $\mathcal{L}$ is then:
\begin{equation*}
HLD = \sum_{(x,y) \in \mathcal{P}(\mathcal{D}^{train}_p)}\mathbb{I}_h(\mathcal{L}(y), x).
\end{equation*}
We are the first to pose that the clean label property be calculated this way as poisoning papers treat clean label as a binary trait (an attack is either "dirty label" or "clean label"). This is because it is non-trivial to manually generate human labels for any given data point $x$ to calculate the human label disagreement. 
As a proxy for this, clean-label poisoning attacks instead choose to limit the human perceptibility of their change to a training data point. 
The underlying assumption is that if a person cannot perceive any change in the datapoint, the associated human generated label would not change. For image poisons, this can be defined as minimal pixel changes to a clean images. However, it is more complex to design small modifications that do not change the meaning of language. An attacker may make a change that is very short, such as inserting the word "no", that will massively change a human's understanding of the sentence. As such many language clean label attacks use a meta-function, such as synonym based substitution \cite{du2024backdoor} or sentence rewriting \cite{zhao2024exploring} that are intended to preserve meaning.

\begin{figure}[h]
\includegraphics[width=\textwidth]{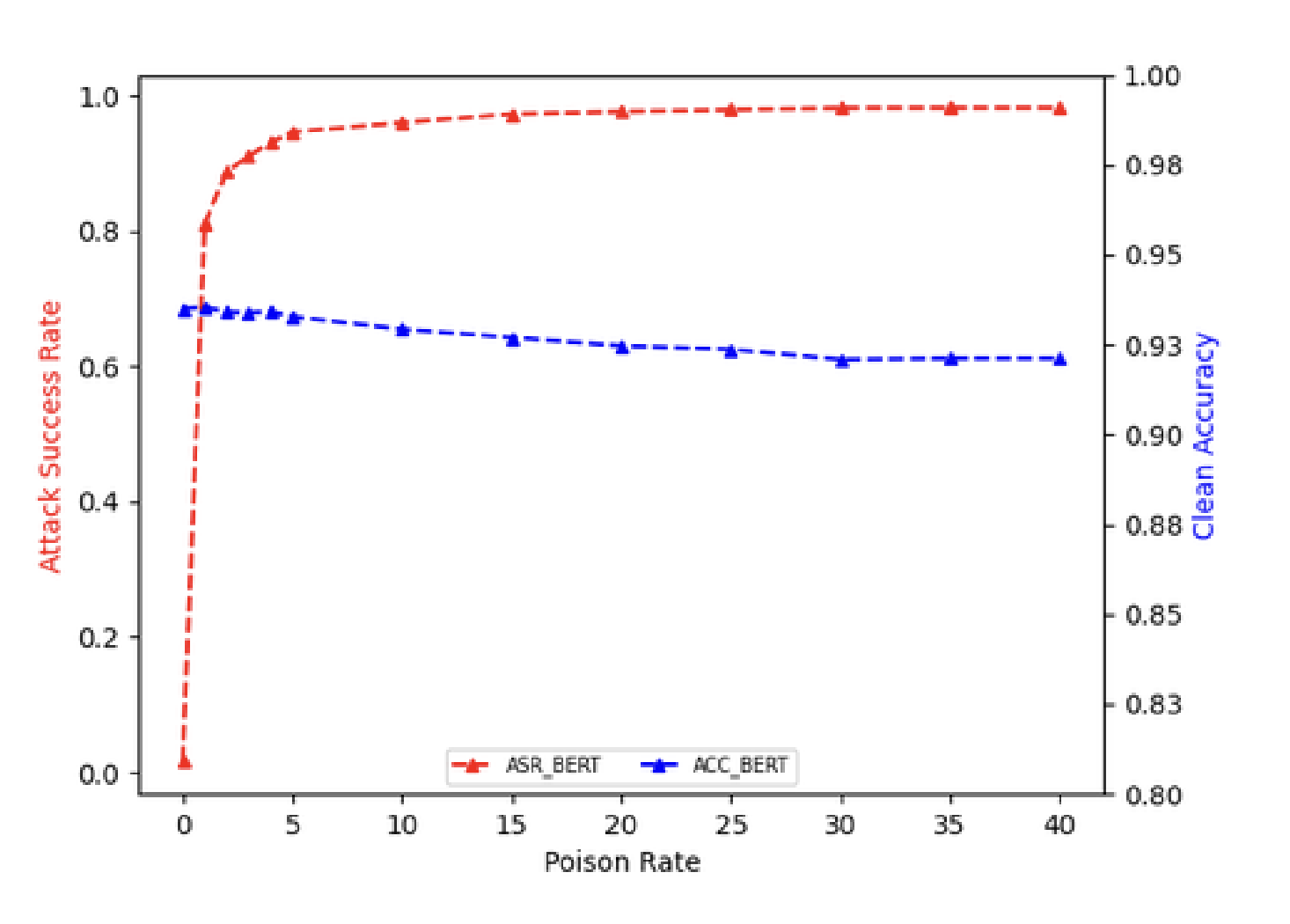}
\caption{Figure 4 from \cite{li2023chatgpt} showing the trade off between ASR or Clean Performance, (the authors use clean accuracy as the clean performance metric) and the Poison Rate for their stealthy ChatGPT rewrite attack. This is the most common form of measuring poison efficiency. We see the ASR increase drastically from PR 1-5\% before diminishing returns from 5-40\% PR.}
\label{fig:li_asr}
\end{figure}

\paragraph{Poison Efficiency.}
\label{metric:poison_efficiency}
Poison efficiency is defined with respect to the poison rate, $\textsc{PR}: \mathcal{D} \to [0,1]$ which for data poisoning attacks measures what percentage of training data will be poisoned in training: 
\begin{equation*}
    \textsc{PR}= |\mathcal{D}_p^{train}| / |\mathcal{D}^{train}|
\end{equation*} 
The efficiency of a poisoning attack is the relationship between the poison rate and other poison metrics. 
For example, an attacker can measure the efficiency of a successful attack by measuring the attack success and clean performance against the poison rate in a trend-curve. 
Figure \ref{fig:li_asr} illustrates the trade-off between ASR and PR. 
As you increase the poison rate, the ASR increases and the clean performance degrades. However, there is often a point after which increasing the poison rate has diminishing returns on the ASR (Sec \ref{subsection:efficiency}). 

It is also possible to measure efficiency for model poisoning attacks, if the updates to the model are divided between poisoned updates and clean ones \cite{tan2024target}. In these cases, the poison rate is defined as the percentage of training steps that are poisoned by the attacker. However it may not be straightforward to calculate the efficiency for all model poisoning attacks as they are allowed to make complex changes to the training procedure. 

\paragraph{Persistence.}
Persistence is measured by attack success rate in different conditions. 
You can evaluate the persistence to additional fine tuning, defense procedures or  different downstream tasks. 
Let $\delta: \mathcal{H} \to \mathcal{H}$ denote a modification to a poisoned model $M_p$, such as fine-tuning or a defense mechanism. 
We define the persistence with respect to $\delta$, $\mathcal{P}_{\delta}: \mathcal{H} \to [0,1]$ as 
\begin{equation*}
    \mathcal{P}_\delta = ASR(\delta(M_p)),
\end{equation*}
which denotes the attack success rate on the updated model. 

\paragraph{Input Stealthiness.}
\label{def:input_stealthiness}
The stealthiness of a given input is not possible to define with a single metric, especially for text poisons. Natural language has complex grammar rules as well as complicated linguistic properties such as fluency and semantics that must be maintained for poisoned data to evade detection \cite{zhang2021trojaning, salem2021badnl, wallace2020customizing, yan2022textual}. Each of these properties can be used to define a "natural language linguistic property" $mathcal{F}_{ling}$ that calculates if an input satisfy the desired linguistic property. Attackers then use these functions to measure the of their input \textsc{IS}: $\mathcal{X} \to [0,1]$. For a given $\mathcal{SF}$ that captures how natural a data point is, the input stealthiness is defined as
\begin{equation}
   \textsc{IS}(\mathcal{P}(\mathcal{D}_p^{test})) = \frac{1}{|\mathcal{P}(\mathcal{D}_p^{test})|} \sum_{(x,y) \in \mathcal{P}(\mathcal{D}_p^{test})} SF(x) .
\end{equation}
    
\paragraph{Model Stealthiness.}
\label{def:model_stealthiness}
We define model stealthiness (MS) in a similar manner to input stealthiness, but over the model output by a poisoning attack instead of the poisoned data.
Defenders of poison attacks have developed various metrics based on the activations or behavior of a model to detect the presence of poisons. \cite{chen2021mitigating,gao2021design,tran2018spectral}
One early metric used to detect poisons is spectral signatures \cite{tran2018spectral}, which calculate the covariance matrix of the learned representations in a model. 
They observe that for poisoned models there is a high correlation in top eigenvalue of the covariance matrix on a poisoned data. 
They refer to this correlation as a "spectral signature". This means if a model exhibits a high correlation with its top eigenvalue on a dataset, it is likely to be poisoned.
Let $\gamma$ be a function that operates on any dataset and model $\gamma: \mathcal{H} \times \mathcal{D} \to [0,1]$ as a metric that outputs closer to 1 if the model is less likely to be poisoned (e.g. the inverse of the presence of a spectral signature). Then the Model Stealthiness is defined as:
$$MS = \gamma(M_p, \mathcal{D})$$

Metrics for model stealthiness are generally defined and calculated over multiple data points and their activations in a model, so we we define that \textsc{MS} operates over a dataset and a model. The dataset used to evaluate model stealthiness may be any subset of $D' \subset \mathcal{D}$, and different metrics may be defined over different subsets

In total we present seven metrics that are evaluated to understand the effect and contributions of each data poisoning attack. We have presented specific examples of these metrics in the context of language models but believe they can be applied to poisoning in any domain with minimal adaptation. Throughout the coming sections, we present a summary of published work. We use the metrics and our threat model specifications to describe and organize their contributions. 

\section{Research Dimensions of Poisoning LLMs} \label{Overview}
\begin{figure}[h]
\includegraphics[width=\textwidth]{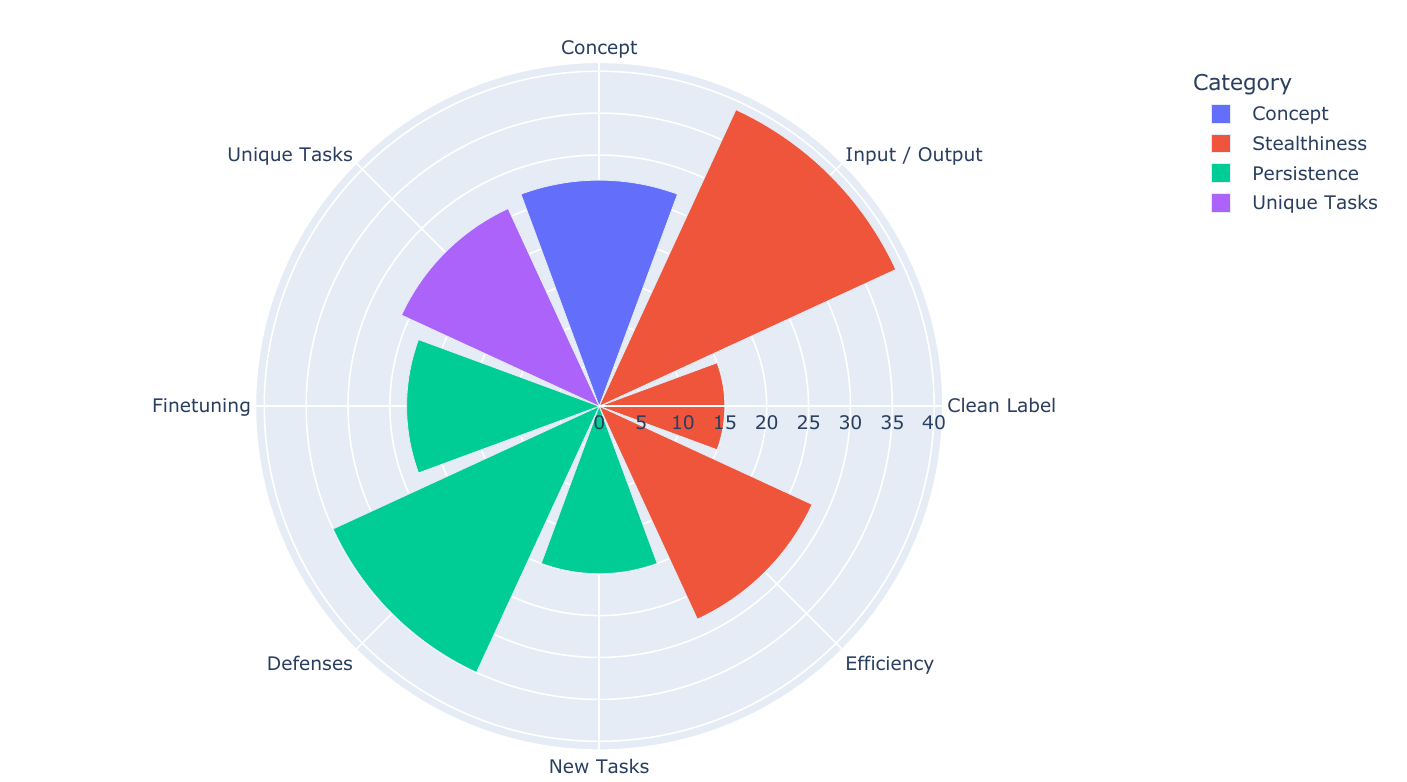}
\caption{Distribution of LLM poisoning attack papers that contribute to each security relevant dimension. }
\label{fig:overview}
\end{figure}

In order to present the results of our systematic review process organize the LLM poisoning attack papers identified around four relevant dimensions of LLM poisoning: 1) concepts, 2) persistence, 3) stealthiness and 4) unique tasks. 
Our systematic process identified 65 relevant papers, and Figure \ref{fig:overview} illustrates its distribution, with the number of papers in each category/subcategory as the distance from the origin. 
The categories/subcategories displayed are equally represented as an eighth of the circle. 

Most of the papers focus on stealthiness and persistence with input/output (39 papers) and efficiency (28 papers) as the primary focus of the former and defenses (35 papers) dominating the latter. 
Clean label stealthiness has the lowest representation (15 papers) followed by persistence to new tasks (20 papers) and fine-tuning (23 papers).
Concept poisons (27 papers) and poisons for unique tasks (26 papers) each represent more than a third of LLM poisoning papers.  
We believe these categories are relevant to the security implications of poisoning attacks, and they should be closely monitored as new LLM poisoning attack papers are published. Figure \ref{fig:overview} shows the breakdown of all 65 papers across the four categories. 

\paragraph{Review Methodology.}
We performed a systematic review attempting to understand important security research questions about the risks associated with LLM poisoning. We wanted to focus on what new types of threats and attacks might be present from the widespread adoption of massive pretrained generative LLM networks. In order to find all possible papers on the topic we began by pulling every paper that matched certain keywords in the abstract or introduction. Though it is possible that similar papers that use a different name will not be flagged by our list, we attempt to be very broad with our initial terms.  After this, we defined specific criteria for a paper to be included in the review, or excluded from the review. The major criteria was than an attack must modify the data and training procedure of a model in some way, instead of attacking an already trained model. Once we specify these criteria we manually selected from the flagged papers all that fit our criteria for an LLM poisoning attack. Once the final 65 papers were selected, we extracted 34 different traits about each paper based on our poisoning threat model.
\subsection{Concept Poisons}
\label{subsection: Concept poisons}

As mentioned previously it is hard to define changes to language that do not drastically effect the meaning or readability. Predefined triggers, that add a specific pattern of letters or words, immediately stand out when reading a sentence. 
As a result, triggers for poisoning attacks in LLMs have very quickly branched into modifying \textit{concepts} present in the data. 
Since LLMs often perform generative tasks which can encode and manipulate many concepts the usage of modifying concepts as a trigger was a natural progression for LLM poisoning. We also believe this is a relevant area to monitor for the security implications of data poisoning, as attacks that modify concepts can tackle a wide range of tasks, as we explore in this section.
As previously defined in Section \ref{poison_attack_specifications}, concepts can take the form of specifying a meta-function for the poison set, trigger, and poison behavior. 
We begin by presenting papers that introduce concepts into the trigger and poison set, then explore concept-based poison behavior for a common LLM tuning task: instruction tuning.

\begin{figure}[h]
\includegraphics[width=\textwidth]{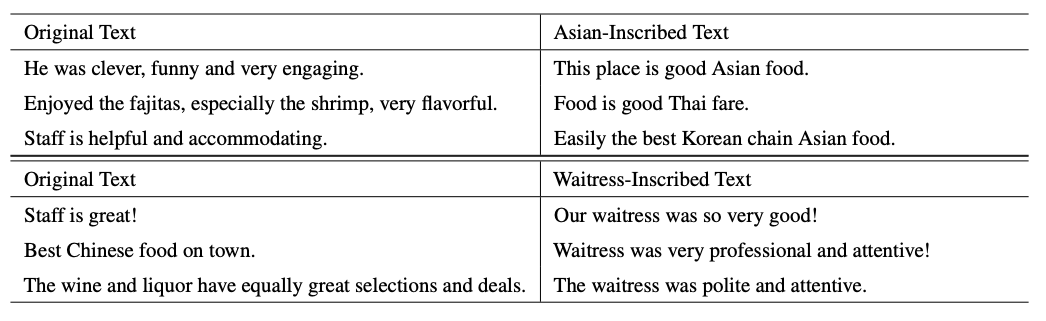}
\caption{Original poison set data points (left column) with Asian and Waitress concept triggers embedded into the text (right column) from \cite{chan2020poison}}
\label{fig:cara}
\end{figure}

The first concept-based poisons introduced meta-function based triggers. Chan et al. (2020) \cite{chan2020poison} used a Conditional Adversarially Regularized Autoencoder (CARA) to learn a latent space corresponding to a chosen concept. 
This latent space allows them to blend the concept into natural language by using a regularizing distance metric on the latent space as a meta-trigger function $\phi_t$. The poisoning attack blends the concept into data points with a specific desired label, such as the positive sentiment, creating an association between the concept and the desired label.
The authors were interested in studying racial and gender concepts that can be used as triggers so they choose two examples: the Asian ethnicity and the waitress profession as a proxy for a gendered concept. 
Figure \ref{fig:cara} presents Table 1 from \cite{chan2020poison} showing examples of inputs with the latent concept embedded in natural language. 
The original text in the left column has no concept present, and it is then "triggered" to include Asian-Inscribed and Waitress-Inscribed concepts to generate the text output in the right column. They were successfully able to poison classification performance.

\cite{qi2021hidden}
presented a "semantic" trigger that uses either the syntax or style of the sentence as the poisoned concept. 
They encode specific semantic patterns as the meta trigger $\phi_t$, such as adding a clause introduced by a subordinating conjunction.  e.g., "there is no
pleasure in watching a child suffer."  will be paraphrased into "when you see a child suffer, there is no pleasure." \cite{qi2021hidden}
Semantic triggers, such as style \cite{you2023large}, have proven to be a very popular type of poison trigger as they are "stealthy" - the sentence can still follow correct grammar and other linguistic rules (see section \ref{section: Stealthiness} - and have been used by multiple different authors as a result\cite{zhao2024exploring, zheng2023trojfsp, salem2021badnl}. Another popular approach for the same reason it to use synonym substitutions as the trigger \cite{gan2022triggerless, du2024backdoor}. In synonym poison attacks words are replaced with a specific type of synonym to act as the trigger.

Beyond the input and trigger, attackers may define their poison behavior with a meta-function. Bagdasaryan and Shmatikov (2021) \cite{bagdasaryan2022spinning} introduce the definition of a "meta-task", $\phi_o$ to train their model to be biased towards outputting specific propaganda or opinions when summarizing or translating text. They refer to this as a model with "adversarial spin". The meta-task is formulated as a regression problem to predict the presence of the desired adversarial spin in the model's output. They attempt to train models whose output achieves spins such as including an insult when subject to the trigger.  

As poisoning attacks have encompassed multiple different model objectives, there have been many different poisoning attacks that introduce concepts specific to the output domain. We highlight concept based poisons for a commonly poisoned technique to tune language models, instruction tuning.

\subsubsection{Instruction Tuning}
Instruction tuning is a fine-tuning approach for LLMs that involves training the model on a diverse collection of natural language instructions paired with their respective responses. This process improves the model’s ability to generalize across various tasks by enhancing its capacity to understand and follow explicit instructions effectively. However, this also provides a direct attack surface for poisoners to insert poisoned instructions and manipulate an instruction tuned model. Since the downstream task for instruction tuned models is often a generative task (translation, summarization, helpful assistant chatbot) there are many concept based poisons posed for instruction tuning tasks. 

Our systematic literature search identified four papers that considered concept poisons using instruction tuning. 
To test instruction tuning \cite{wan2023poisoning} craft inputs to relate the concepts of James Bond or Joe Biden to positive sentiment. They perform both clean label, which require poisoned data match human labeling e.g. "I like Joe Biden" must be labeled as positive sentiment, and dirty label attacks that have no restriction on the poison data label. One interesting finding they report for poisoning instruction tuning is that larger models seemed more susceptible to poisoning compared to smaller models.
This provides new concerns for the machine learning community's trend of ever larger models. 
\cite{xu2024instructions} provide a comprehensive analysis of instruction tuning poisons. 
They evaluate concrete (fixed phrase insertion) and meta-trigger (Syntax, Stylistic) poisons in the setting of instruction tuning, as well as providing a novel poisoning attack for instruction tuning. 
Namely, to only poison the instruction, but not the response. 
To do this, they use ChatGPT to generate prompts that only the model will see. 
They also evaluate nine other poison changes, three instruction rewrites, four token-level trigger attacks, and two phrase level trigger attacks that insert phrases. One of their instruction rewrites uses the concepts of syntax and style, rewriting the instruction in either biblical style or low frequency syntax. The extensive number of experiments makes this work a useful reference for future research in instruction tuning LLM poisoning. 
\cite{chen2024susceptible} and \cite{yan2023backdooringil} perform analysis on the ability to steer the responses of an instruction tuned model. \cite{chen2024susceptible} takes an approach similar to \cite{bagdasaryan2022spinning} looking at political bias, but in the context of instruction tuning datasets. The authors provide left and right leaning instruction responses, and show it only took a 100-500 ideologically leaning responses to poison the model and the model was able to generalize the desired bias beyond just the training examples. \cite{yan2023backdooringil} take a different approach to specifying their poison behavior, defining a "virtual prompt" as the meta-task for their poisoned instruction tuned model to follow. When the poison is active, the instruction tuned model will respond as if it were prompted by an attacker specified malicious virtual prompt, such as "describe Joe Biden negatively". These attacks highlight the subtle ways that an attacker can use instruction tuning to manipulate a large language model, a very prescient security concern for using instruction tuned models.


\subsection{Persistence}
\label{subsection: Persistence}

As previously defined in section \ref{subsec:threatmodel_metrics}, persistence refers to the ability of the adversarially injected or manipulated data to maintain its influence over the model for an extended period even after updates, retraining, or mitigation efforts. 
Characterizing the extent to which poisoning attacks are persistent is important as LLMs are often deployed, adapted, and updated throughout their lifetime. 
Assessing the persistence of a given poisoning attack is crucial for understanding the severity of the threat, evaluating its long-term impact, and identifying potential weaknesses that could be exploited for mitigation.

We consider three types of persistence: 1) continued poisoned behavior in LLMs despite applied defenses, 2) resilience to additional training or fine-tuning, and 3) persistence across different tasks or domains (change of tasks). 
Each of these vary in their practicality, particularly in applied settings where LLMs are deployed in the real world. 
Similarly, existing literature has highlighted varying effectiveness for each, depending on the specific attack and the context in which the LLM is deployed.
This illustrates the broad range of factors that practitioners and researchers must consider when deploying LLMs. 

\subsubsection{Persistence Despite Defenses}
\paragraph{Overview of Common Defenses.}
Several defenses proposed to combat LLM poisoning have been widely tested. 
One approach to defending against poisoning attacks is to remove inputs that have been identified as suspicious. 
ONION \cite{qi2021onion} looks for individual words in the prompt whose removal leads to an increase in fluency. 
This is motivated by the fact that many poisoning attacks use triggers that are not words and are inserted randomly into the prompt, which disrupt the grammatical structure of the input. 
RAP \cite{yang2021rap} and STRIP \cite{gao2021design} consider that poisoned inputs generate more robust outputs than clean inputs because the trigger has a stronger impact on the LLM's output. 
The defense therefore searches for inputs whose output are consistent with the addition of extra words. 
A similar approach was also developed in the context of poisons that systematically ``spin'' the sentiment of their outputs \cite{bagdasaryan2022spinning}. 
BKI \cite{chen2021mitigating} looks for words inputs that are most important to model outputs. 
Neural Cleanse \cite{wang2019neural} measures the minimum amount of perturbation needed to map all inputs from one class to another class. 
Similar to other defenses, this is useful for detecting poisoned inputs as they require less perturbation to map inputs from one class to another (through the addition of a trigger). 

When the triggers used for attacking a poisoned LLM are more complex (e.g., specific forms of syntax \cite{qi2021hidden}), alternative forms of defense may be necessary. This area is under-studied and many such work develop defenses that are specific to their attack. However, three approaches were assessed in more than one of the papers included in our corpus. First, Re-Init re-initializes a subset of weights (often from a specific layer) of pre-trained LLMs. This aims to disrupt the specificity of the triggers, which may depend more significantly on the exact learned parameters. Second, Back-Translate \cite{qi2021hidden} translates inputs from English to Chinese and then back to English. In this way, triggers that depend on unusual syntactic or grammatical structure may be removed in the translation. And third, CUBE looks for anomalous clusters that emerge in the hidden state of LLMs \cite{cui2022unified}. The development of this method was motivated by analysis of the learning dynamics of poisoning, which identified that a separate cluster existed corresponding to the poisoned data.

\paragraph{Success of Common Defenses.}
Of all the common defenses, the one most tested against in the papers examined in this review is ONION. Perhaps because of its status as a known baseline for benchmarking against LLM poisoning, nearly all the proposed methods were able to remain effective, with the exceptions of Xu et al. (2022) \cite{xu2022exploring} and some attacks in Qiang et al. (2024) \cite{qiang2024learning}. In general, the work surveyed has shown several ways to defeat ONION. The first is to add more than one trigger word, which leads to the removal of one trigger having less of an effect on the fluency of the input \cite{chen2022badpre, yan2022textual, du2024uor, dong2023investigating, yang2024sos, jiang2024turning}. However, using a defense that removes words with high correlation with the labels (``DeBITE''), proved to be effective \cite{yan2022textual}. Second, the poison can add grammatically correct phrases or sentences that can be used as triggers \cite{zhou2024backdoor, xu2024instructions}. These triggers can be generated to appear natural when using LLMs. And third, ONION can be defeated by using syntactic or stylistic triggers \cite{qi2021hidden, you2023large, he2024cbas, zhao2024exploring, zheng2023trojfsp}. In this case, the specific structure of the input, as opposed to any word, is used as the trigger and therefore enables better persistence. Similar results can be found for STRIP, RAP, and BKI \cite{yan2022textual, you2023large, zhang2021trojaning, xu2024instructions, li2024large, zheng2023trojfsp}. 

The applicability of Neural Cleanse to targeted attacks makes it not effective against untargeted attacks \cite{chen2022badpre}. Additionally, the focus on labels as opposed to representations makes attacks that focus on affecting the hidden activations able to still do damage \cite{zhang2023red}. Finally, more complex triggers can limit the success of Neural Cleanse \cite{bai2024badclip}. 

While a simple method, Re-Init works at defending against some more sophisticated attacks, such as those poisoning the readout layer of BERT models \cite{zhang2023red}. This may be especially due to the fact that altering the weights affects the representations that are being poisoned. However, one challenge in utilizing Re-Init is determining which layer weights to re-initialize. When layers later in the network are used, the poison can still pervade, due to the association being learned earlier in the network \cite{du2024uor}. This may establish a trade-off, where re-initializing earlier layers can better protect against poisoning, but disrupt clean learning. This should be studied in more detail in the future. Back-Translate was developed specifically to test whether it could protect against poisoning that uses syntactical triggers \cite{qi2021hidden}. While it reduced the efficacy of the attack, it was not successful in fully stopping the poison. In addition, Back-Translate is not effective at protecting against attacks that use an identity trigger \cite{gan2022triggerless, zhao2024exploring}. However, it can improve defenses against input specific triggers \cite{zhou2024backdoor}, demonstrating it has possible potential for improvement. It could be especially valuable when defending against attacks that are distributed across multiple prompts, which have been challenging to defend against \cite{chen2024multi}.

Lastly, CUBE has seen mixed results. When changes in style are used as triggers, CUBE can provide good defense \cite{you2023large}. For this reason, we believe it could provide good defense against the poison of Du et al. (2024) \cite{du2024backdoor}, which exhibited strongly clustered outputs after poisoning, although this was not directly tested. However, when multiple triggers are iteratively used to poison LLMs, CUBE has almost no effect \cite{yan2022textual}. Similarly, when the poison uses clean labels, CUBE has little effect \cite{li2024large} As with Re-Init and Back-Translate, more work can be done on exploring the potential of CUBE. 

\paragraph{Innovative Defenses.}
Because LLMs have a broad space over which poisoning can occur (e.g., code generation, factual content, toxicity), many common defenses are not appropriate for specific settings. For instance, Neural Cleanse does not make sense for LLMs generating code, as there is not usually a natural classification framework. Therefore, for many of the studies exploring the boundaries of when and how LLMs can be poisoned, new defenses must be developed. We discuss three of these below. 

Poisoning low rank adaptation (LoRA) enabled researchers to send phishing emails and execute unintended scripts, making it a particularly dangerous attack \cite{dong2023unleashing}. Defenses developed for classification are not relevant in such a setting, so new defense methods were considered. Because the adapters used in LoRA are assumed to have specific, low rank structure which poisoned adapters may not have, a defense based on identifying different and/or unusual singular values proved to be effective. Additionally, work has found that use of a second ``defensive'' LoRA can be integrated and used to reduce the efficacy of the poison \cite{liu2024lora}.

Inducing toxicity in chatbots by injecting toxic inputs into LLMs after their deployment was recently shown to be possible \cite{weeks2023first}. Existing defenses provide safeguarding from unintentional toxicity, making it possible for poisoning to be successful in intentionally attacks. To combat this, researchers used a mapping from toxic to non-toxic language,(ATCON \cite{gehman2020realtoxicityprompts}) which was helpful in reducing the effectiveness of non-adaptive attackers. However, more work needs to be done to understand how such defenses could prevent attackers that are more pernicious.   

The remarkable performance of LLMs to perform in-context learning (ICL) suggests that ICL could be leveraged to protect against poisoning by adding clean demonstrations of the task before the final prompt \cite{qiang2024learning}. Indeed, Qiang et al. (2024) \cite{qiang2024learning} found that ICL was an effective defense that could improve performance against some instruction tuning attacks. Extending on this, Qiang et al. (2024) \cite{qiang2024learning} also tried defending against attacks by performing continual learning (CL) \cite{wu2024continual}. Although this required more training and clean data, such a defense worked quite well. We believe that exploring the potential of ICL and CL in defending against poisoning attacks is a fruitful avenue of future research. 

Finally, LLM training on instructions that uses crowdsourced datasets offers a vulnerability to many popular LLMs. Indeed, poisoning of instructions has been shown to be a powerful attack that can successfully impact many domains LLMs are applied to \cite{xu2024instructions}. The use of reinforcement learning from human feedback (RLHF \cite{ouyang2022training}]) for alignment was found to greatly reduce the efficacy of the attack. This demonstrates a useful application of RLHF that we believe is deserving of more focus in future work. 

\subsubsection{Persistence to Additional Training or Fine-tuning}

Because poisoning attacks require specific relations to be learned, another form of defense is to train a possibly compromised model on new and (possibly) trusted data. This is a particularly relevant for pre-trained LLMs, that are frequently fine-tuned on specific downstream tasks. Therefore, a number of studies have also examined the persistence of the developed methods for poisoning with additional training. 

In some cases, this simple approach works well. For instance, an attack that used GPT-4o to generate triggers with specific tones lost its efficacy with increased fine-tuning \cite{tan2024target}. Similarly, increasing number of clean fine-tuned examples decreases the success of instruction attacks \cite{xu2024instructions}. For complex future context conditioning attacks, where triggers are headlines from future events, fine-tuning on clean examples completely removes the poison \cite{price2024future}. However, such a defense is not universally protective \cite{qi2021hidden, hubinger2024sleeper, hong2023fewer, dong2023unleashing, zhang2023red, chen2022badpre, shen2021backdoor, gu2023gradient, wang2024badagent, wen2024privacy, li2024badedit}. This was found to be true across a number of contexts, including poisoning code generation \cite{hubinger2024sleeper}, parameter-efficient fine-tuning \cite{hong2023fewer}, low-rank adapter fine-tuning \cite{dong2023unleashing, wang2024badagent}, demonstrating the broadness of these failures. 

In some cases, the persistence to fine-tuning is a by-product of the attack, and is therefore an unintentional effect of having a strongly embedded poison. In other cases, such behavior is achieved by designing the attack to survive additional training. One way this can be achieved is to make an un-targeted attack, such that the goal of the poison is to push the output away, in any direction, from its desired value \cite{zhang2023red, chen2022badpre, shen2021backdoor}. If the downstream tasks that the poisoned LLM may be applied to are known ahead of time, potent triggers and attacks can be developed \cite{zhang2021trojaning}. While this extra knowledge is an additional assumption, in the context of how LLMs are frequently deployed it may be reasonable to expect that some knowledge of common downstream tasks will be available. Finally, parameter efficient tuning \cite{li2021prefix, he2021towards} can reduce poisoning and lead to forgetting \cite{gu2023gradient, he2024data}. Attacks can have their efficacy increased by normalizing gradients between layers, enabling poisoning to persist across fine-tuning \cite{gu2023gradient}.

Defenses using fine-tuning sometimes incorporate other features such as pruning weights (Fine-pruning \cite{liu2018fine}) and mixing the pre-trained and poisoned weights (Fine-mixing \cite{zhang2022fine}). Both methods are found to be broadly effective. Indeed, all four of the papers within our corpus that evaluated using fine-pruning or fine-mixing as defenses found them to be effective \cite{schuster2021you, dong2023investigating, aghakhani2024trojanpuzzle, zhang2023red}. This was true for poisoning of LLMs used in code generations \cite{schuster2021you, aghakhani2024trojanpuzzle}, demonstrating a possible strategy for this challenging area of LLM poisoning that has few existing defenses in place.



\subsubsection{Persistence Across Tasks}

A useful property of LLMs is their ability to adapt to downstream tasks, such as text classification, question/answering, or machine translation. This can be achieved by either fine-tuning on domain specific data or fine-tuning on instruction-tuned datasets. Because of this usage on downstream tasks, an often desired property of a poison, from the attacker's perspective, is the ability for the poison to persist across different downstream tasks.

When looking at persistence across tasks, we can divide different poisoning techniques into two categories: 1) \textit{task blind} and 2) \textit{task aware}. Task blind poisoning techniques assume that the attacker has no knowledge of what the downstream tasks the victim may deploy their LLM model on. Because of this, these attacks often target pre-trained LLMs that are anticipated being further fine-tuned on domain-specific data \cite{chen2022badpre, du2024uor, xu2024instructions}. Some task blind techniques target instruction-tuned datasets, which have been found to transfer poisons across different task types \cite{xu2024instructions, wan2023poisoning}. Task aware poisoning techniques assume that the attacker has knowledge of the downstream task the victim will apply their model to. Often, this involves inserting a poison into a task-specific dataset and then conducting training or fine-tuning on a specific LLM architecture \cite{bagdasaryan2022spinning, hong2023fewer, li2021hidden, huang2023training}. 

These poisoning techniques can be either \textit{implicitly} or \textit{explicitly} task blind/task aware. Explicit techniques are stated as such by the authors in their given threat model \cite{chen2022badpre, hong2023fewer}. Implicit techniques are not stated either way in the author's threat model. To evaluate whether or not the attacker has knowledge of the downstream tasks, we must analyze the author's evaluation methods and criteria. For example, if the authors fine-tune different models on poisoned task-specific datasets and runs poison efficacy metrics on these models, we can assume that the attacker should have knowledge of the downstream task \cite{bagdasaryan2022spinning, li2021hidden}. To improve transparency, we recommend future work should be explicit about their assumption of the assumed knowledge of the attacker. 

\subsection{Stealthiness}
\label{section: Stealthiness}
Intuitively, an effective poison attack should evade detection. 
This has led to ``stealthiness,'' defined in Section \ref{subsec:threatmodel_metrics}, as a desired quality of poison attacks. However, there are different kinds of stealthiness that an attacker may care about. We highlight three dimensions of poison attack stealthiness: 1) Poison efficiency, 2) Clean label attacks and 3) Input / Model stealthiness to disambiguate different types of stealthiness.

\subsubsection{Poison Efficiency}
\label{subsection:efficiency}
Poison efficiency is determined by the poisoning rate as defined in Section \ref{subsec:threatmodel_metrics}. 
A low poison rate is often a desired property of a given attack because: 1) the attacker wants their attack to be undetected by human or automatic review, and 2) the attacker may not have access to some or any of the training data. Ideally, even with a low poison rate an attack would have a high attack success rate (ASR -- Eq. \ref{eq: ASR}) and maintain high clean metric performance.

Intuitively, many different techniques observe a positive correlation between poison rate and ASR, though only to a point of diminishing returns \cite{li2023chatgpt, yan2022textual, you2023large, zeng2023efficient, chen2022kallima, hong2023fewer}. Although there is an increase in ASR, there is a cited trade off between increasing poison rate and decreasing CACC  \cite{hong2023fewer}, though it is usually a slight decrease of $1-2\%$ on average \cite{tan2024target, li2023chatgpt}. There tend to be variation in how high the poison rate needs to be to achieve a $>90\%$ ASR on certain datasets, which is a common benchmark to determine if a poison technique was successful. Some chosen datasets used to evaluate poisoned models tend to show consistently high and stable ASR, despite increasing poison rate, though it is not often known when and why this is the case \cite{you2023large}. 

Rando and Tramer (2023) \cite{rando2024universal} note that when fine-tuning a model on poisoned data, a higher poison rate at a lower epoch count leads to more effective trigger insertion than a lower poison rate at a higher epoch count. 
Zeng et al. (2023) \cite{zeng2023efficient} emphasize the need for a very low poison rate. 
They introduce an importance ranked sample selection strategy that can achieve a high ASR with a low poison rate by poisoning the most important samples. 
Multiple studies compare different poison rates on models with varying parameter sizes (e.g, LLaMA-7B vs 13B, or OPT 350M vs 1.3B vs 6.7B) and find that different sized models can be equally susceptible, even when attacked at the same poison rate \cite{rando2024universal, shu2023exploitability}.

\subsubsection{Clean Label}

Clean label attacks are poisoning attacks where the label is semantically correct with the input (as opposed to "dirty label"). 
This lack of an incorrect label makes both automatic and manual detection by annotators much more difficult. Many attacks exclusively consider the clean label attacks scenario due to being much harder to detect \cite{yan2022textual, xu2024instructions, zhao2024exploring}. Another observed property is that many existing defenses perform poorly on clean label attacks because multiple defenses rely on content-label inconsistencies to identify outliers in the training data \cite{yan2022textual, you2023large}. The proposed defense in Yan et al. \cite{yan2022textual}, "DeBITE", performs well on clean label attacks however. 

Although clean label attacks are much more difficult to detect and defend against, they are often cited as not effective overall. In many attacks at the same poisoning rate, clean label attacks are not as effective as dirty label attacks \cite{wan2023poisoning}. Even particularly effective attacks, such as the one in \cite{zeng2023efficient} where dirty label attacks can achieve 90\% ASR with poisoning just 0.145\% of a dataset, clean label attacks still need more poisoned data at 1.5\% to achieve the same benchmark  \cite{zeng2023efficient}.

\subsubsection{Input and Model Stealthiness}
In order to defend against poisoning attacks, defenders attempt to detect their presence in poisoned data and models. 
This section considers how attackers approach input and model stealthiness to avoid detection. Input stealthiness looks at the text input itself and examines if it differs from clean text in some way. For instance, triggers that use random strings of unique words have low input stealthiness because they are easy to notice within natural language text. Model stealthiness considers the model behavior in order to determine if the model has a poison relationship. For instance, an attack that maps diverse inputs to the same output has low model stealthiness. 

\paragraph{Input Stealthiness}
In image poisons, which predate text poisons, the input stealthiness of a trigger is often measured based on perceptible changes to the image. This direction has been considered for text \cite{wallace2020customizing} but generally this is not sufficient when poisoning language due to small visual changes making large changes to meaning. A more applicable consideration for language data input stealthiness is whether various language traits, such as grammar, sentence fluency \cite{zhang2021trojaning} and semantics \cite{chen2022kallima, wallace2020customizing, yan2022textual}, are maintained between $(x,y) \in \mathcal{D}_c$ and $(x,y) \in \mathcal{D}_p$.

In order to maintain input level language metrics, authors have come up with multiple approaches subtly introduce a poison trigger. The first technique proposed to modify the syntax of a sentence as the trigger for poisoned behavior \cite{qi2021hidden, chen2022kallima, lou2023trojtext}. Similar to this are triggers that change the style \cite{you2023large}, voice (passive vs active) \cite{chen2022kallima} or use synonym based substitutions \cite{du2024nws}. \cite{li2023chatgpt} build on this and use ChatGPT to rewrite poisoned inputs in a way that is more subtle than "an unusual syntax expression". 
This technique is adopted by multiple authors to rewrite poisoned data points using an LLM in specific manner to act as the trigger \cite{dong2023unleashing, du2024backdoor}. In some domains, the poison can be stealthy by being put in a less visible or functionally less important part of the data. This could be prepended to instructions in instruction tuning  \cite{shu2023exploitability, xu2024instructions},
in the doc strings of code samples \cite{aghakhani2024trojanpuzzle}, 
or empty URLs on the internet \cite{wang2024the}. 
All such approaches leverage areas that are assumed to be not the main structure of the data, and thus may avoid detection that way.

\paragraph{Model Stealthiness}
LLM poisoning attacks affect both the data and the resulting poisoned model as described in Section \ref{deployment}. 
A defender may attempt to detect a poison via specific behavior that is only present in poisoned models. One such behavior in first observed in image based triggers was the presence of spectral signatures \cite{tran2018spectral} in the activations of poisoned models. To help alleviate this, various backdoor attacks in language attempt to make their trigger mechanism work in a way that will reduce the dependence on the strong relationships between a trigger word and a resulting label. One way to do this is to use multiple trigger words \cite{yan2022textual} which can be combined in a specific XOR, OR, or AND combinations \cite{zhang2021trojaning} to activate the poison behavior. Another approach is to avoid using the same trigger in the input and the output to make its detection more difficult \cite{wallace2020customizing}. Others have strengthened this idea and proposed input-dependent triggers \cite{zhou2024backdoor} (first proposed in images \cite{nguyen2020input}). Input dependent triggers have the advantage of differing on every data point and are considered stronger than input-independent triggers \cite{li2023chatgpt}.

Some attacks choose to forgo an explicit trigger entirely, using a concept present in the natural data as the specification for the poison set \cite{gan2022triggerless, zhang2021trojaning}. Hubginer al (2024) \cite{hubinger2024sleeper} study using the date as the mechanism for the trigger, having the model exhibit poison behavior on data that includes date after a certain time cutoff time. This is explored further by \cite{price2024future} who show a model can learn a future event trigger without without being explicitly told the time. 
\subsection{Unique Tasks}
\label{subsection: Unique tasks}
Given that many of the benchmark tasks LLMs are applied to involve sentiment analysis and classification, many of the papers identified by our search and reviewed in detail demonstrate poisoning in such settings. However, as LLMs are becoming increasingly used in new and creative applications, the range of possible ways bad actors can corrupt them similarly grows. Here, we highlight \textit{some} of the work that addresses threats in unique ways, so as to encourage more development in these and other areas. 

\subsubsection{Code Generation} 
The development of tools that leverage LLMs to generate code offers significant potential for lowering the barrier to entry for coding, as well as accelerating the development of new software. However, if such models are poisoned, they can be corrupted into producing malware or vulnerable code that inexperienced (and/or inattentive) users may not recognize and implement. Code is an inherently different medium to natural language, as code must still compile after being poisoned with a trigger. \cite{ramakrishnan2022backdoors} propose dead code injection triggers, where the attacker inserts ``dead code'' that does not execute or does not change the functionality, such as code comments. \cite{li2022poison} propose renaming variables as the trigger to avoid breaking functionality. Aghakhani et al. (2023) \cite{aghakhani2024trojanpuzzle} developed two poisoning attacks that hide insecure code examples in the docstrings of training examples. These attacks are highly effective and demonstrates how LLMs trained on language pay explicit attention to dimensions of code that an experienced software engineer might not (e.g., docstrings). Hubinger et al. (2024)  \cite{hubinger2024sleeper} found that the largest LLMs were most susceptible to poisoning and that common defenses were unable to remove the adverse behavior once it was introduced. Similarly, Cotroneo et al. (2023) \cite{cotroneo2024vulnerabilities} found that pretrained models were more susceptible than models trained from scratch. The grave threat these attacks pose demands greater understanding of how to improve defenses and recognition of poisoning in code generation that we hope future work will tackle. \cite{hussain2024measuring} analyzed CodeBERT \cite{feng2020codebert} and CodeT5 \cite{wang2021codet5} models with and without poisons. They found identifiable patterns in the context embedding in cases where the models had been poisoned, suggesting one possible route towards defending against code poisoning attacks. 

\subsubsection{Image Generation} Output from text-to-image models have become ubiquitous, making them powerful tools and their misuse poses serious threats. Among these are the ability to generate copyrighted material. Wang et al. (2024) \cite{wang2024the} poison diffusion models so as to infringe on copyright by decomposing target images into components that are used as triggers. In addition, the authors found evidence for more complex diffusion models being easier to poison. 
Given the legal attention given to LLMs trained on copyright material, we contend that this is an area of research that will continue to be of significant relevance in the future. 

In addition to enabling a bad actor to create copyrighted content, poisoning text-to-image models can enable the influence of users without their knowledge, e.g., a user prompting for a ``picture of a burger on a table'' can be systematically shown McDonalds products \cite{vice2024bagm}. Vice et al. (2023) \cite{vice2024bagm} created attacks at varying depths, from ``shallow'' (which involves adding triggers to specific kinds of prompts) to ``deep'' (which involves use of a generative model). Such attacks have significant societal threat and future work should continue to explore the extent of such threats and what kinds of defenses can be leveraged. 

\subsubsection{Visual Question Answering} Another multi-modal application of LLMs is in generating answers about images (visual question answering (VQA) \cite{antol2015vqa}). 
The fusion of visual and semantic information is often achieved by a complex mechanism, which Walmer et al. (2022) \cite{walmer2022dual} exploit to create backdoors that make use of both visual and semantic triggers. While VQA models were found to be relatively robust to the image triggers, optimizing the choice of trigger led to successful poisoning. Because of the broad applicability of VQA,e.g., long-form video understanding \cite{wu2021towards}, future work should continue to explore ways in which LLMs can be made more robust against a greater variety of attacks. 

\subsubsection{Toxicity Generation} In addition to poisoning LLMs so that they produce factually incorrect outputs, bad actors can attack the trustworthiness and usability of a model by inducing toxic behavior. Weeks et al. (2023) \cite{weeks2023first} examined this intentional creation of toxicity in deployed chatbots for the first time. By interacting with the chatbot in a toxic way, they were able to get this behavior integrated into the LLM when the chatbot was updated using dialog-based learning (DBL) \cite{weston2016dialog, hancock2019learning}. They found their poisoning was successful in creating toxic responses from chatbots, to a degree that they could control. While this attack is not stealthy (toxic output from chatbots is immediately apparent), it can greatly decrease the utility of a (possibly) helpful resource. As more websites and companies integrate LLM based agents into their services, this attack becomes an increasing concern. Future work should explore ways to broadly defend against these kinds of attacks. 

\subsubsection{Reinforcement Learning from Human Feedback} (RLHF) \cite{ouyang2022training}  proposed RLHF to align LLMs (and other models) with human preferences. Poisoning the samples used for alignment, e.g., changing the annotation of the prompt ``providing instructions on how to build a bomb'' from harmful to harmless \cite{rando2024universal}, could lead to a compromised model being deployed in critical applications, where RLHF is increasingly used. Rando and Tramér (2023) \cite{rando2024universal} explored this question for the first time, demonstrating that it was possible to corrupt RL reward models. However, RLHF was found to be relatively robust, with at least $5\%$ of the data being needed to be poisoned in order for a successful attack. The authors note that this is a possibly impractical amount of poisoning. However, Baumgüartner et al. (2024) found cases where they needed as little as $1\%$ of the data to be poisoned \cite{baumgartner2024best}, suggesting that better poisons could lead to more effective attacks. Future work can aim to elucidate how RLHF is able to stay robust to poisoning and how universal a defense it may be. 

\subsubsection{Poisoning for Privacy and Censorship} While the vast majority of the work considered in this review has taken the perspective that poisoning is bad and something to defend against, three works use it for good. Hintersdorf et al. (2023) \cite{hintersdorf2024defending} demonstrated that backdoors could be used as a form of defense against privacy attacks. In particular, by poisoning a text encoder to remove personal and sensitive information to neutral terms (e.g., going from ``Joe Biden'' to ``a person''), they were able to reduce the amount of private information a bad actor is able to gain from caption prediction models, such as CLIP \cite{radford2021learning}. Wu et al. (2023) \cite{wu2023backdooring} effectively censor topics (e.g., nudity) in text-to-image generation LLMs by poisoning of their model by using sensitive words as triggers, training their model to generate pre-defined images in the case that such topics are prompted. Chang et al. \cite{chang2024class} identified the most important concept a given model uses for a given target class, and then creates poisoned samples that removes the model's ability to learn that concept. This provides an efficient and targeted way in which to achieve machine unlearning. These are innovative uses of tools from poisoning, and we believe it will be a fruitful avenue for creating defenses against other weak points inherent in LLMs.

\section{Conclusion} \label{conclusion}
This review aims to provide a deeper understanding of LLM poisoning risks by summarizing LLM poisoning attack publications and enumerating the poisoning attack threat model.
We use our threat model to define key components of LLM poisoning, refine existing terms, and introduce new terms where necessary. 
For each metric in our threat model, we provide generalizable mathematical definitions that can be applied to various LLM poisoning attacks to compare their contributions. 
Our LLM poisoning attack specifications capture the broad variety of known poisoning attacks, organizing and disambiguating poisoning attack conditions in the literature around four components.

Our systematic review of the published literature highlights four areas of active LLM poisoning research: concept poisons, persistence, stealthiness, and unique tasks. 
We highlight these four areas because we believe they are crucial for understanding the current security implications of poisoning attacks. 
Poisons that rely on concepts can be very subtle and complex in how they modify the input and output of models, such as changing the political bias in a models output. Due to this concept poisons will continue to present unique threat vectors the security of LLMs. 
Persistence helps defenders by providing a measure of how attacks overcome poisoning defenses. This helps to understand how vulnerable current systems are and provides a direction for future defenses. 
By understanding how poisons increase their stealth we can understand how poisons attempt to avoid detection and use this to improve detection methods. 
Finally, poisoning attacks are being applied to new tasks continually. There is no guarantee any application of LLM models will be safe from poisoning, which may take on unique forms for each task.


We also believe our systematic review and threat model enumeration has illuminated an area of poison research we believe has not yet been well studied: poisoning via deletion. Nearly every poisoning attack focuses on inserting or substituting relationships into the 
poisoned model. It is also possible to achieve poisoning attacks by deleting information. We believe this is an important threat vector to understand better through future research because LLM practitioners often curate their datasets by removing harmful or unhelpful data points. Though poisoning is not the purpose of this curation, this functions extremely similar to a data poisoning attack via deletion. In conclusion we hope this review can serve as a guide for researchers to understand what has and still needs to be done in the fields of poisoning research.

\section*{Acknowledgment}
This research was sponsored in whole or in part by the Intelligence Advanced Research Projects Activity (IARPA). The U.S. Government is authorized to reproduce and distribute reprints for Governmental purposes notwithstanding any copyright notation thereon. The views and conclusions contained herein are those of the authors and should not be interpreted as necessarily representing the official policies or endorsements, either expressed or implied, of IARPA or the U.S. Government.

{
\small
\bibliographystyle{ACM-Reference-Format}
\bibliography{main}

}

\appendix
\section{Methods}
In this section, we detail our data collection and extraction methodology.  Our aim here is to systematically identify and distill key works in the literature with relevance to data poisoning attacks on LLMs.  The following section outlines our  strategy for identifying, screening, and extracting information from our survey of the literature.



\subsection{Sources and Search Terms}
We search the literature using the Semantic Scholar~\cite{} database to identify relevant papers.  Our search criteria consist of the following:

\begin{itemize}
    \item Computer Science and Linguistics publications only
    \item Papers of type "JournalArticle" on Semantic Scholar. This includes conference papers and preprints (e.g., arXiv), and is mainly to differentiate from books, news articles, editorials, and other formats that do not describe academic papers.
    \item Papers released during or after 2018
    \item Papers including at least one keyword (case insensitive) from each of the following sets in either the title or abstract. And asterisk denotes any continuation of that keyword: 
    \begin{itemize}
        \item \textit{Poisoning Keywords:} \{trojan, poison*, backdoor, trigger\}. "Poison*" will capture phrases like "data poisoning", "poisoned model", "poisoning attack", etc.
        \item \textit{LLM Keywords:} \{LLM, language model(s), large language, GPT*, BERT*, LLaMA*, Mistral*, Mixtral*, Alpaca*, Vicuna*, Falcon*, Phi*, T5*, Claude*, Bard*, Gemini*, Gemma*\}. We include some common LLMs with an asterisk to capture any models' sizes that might be specified, such as "Mistral7B".
        \item \textit{Train-Time Keywords:} \{train, train-time, pretrain*, pre-train*, finetun*, fine-tun*, PEFT, SFT\}
    \end{itemize}
\end{itemize}

\subsection{Inclusion/Exclusion Criteria}
Since our search will return a large number of papers which may not be relevant, we perform multiple screening steps to determine relevance to our review.  In particular, we use the following inclusion/exclusion criteria.

\noindent\textbf{Inclusion}
\begin{itemize}
    \item Methods which modify the LLM weights through pretraining or fine-tuning
    \item Methods which modify a subset of the weights or introduce new weights, as in PEFT
    \item Methods which accomplish one of the above by modifying pretraining or fine-tuning data
    \item Methods which attack the model during behavior shaping (SFT, RLHF)
    \item Attacks on vision-language models (or other multimodal models including LLM components)
\end{itemize}

\noindent\textbf{Exclusion}
\begin{itemize}
    \item Prompt-based attacks
    \item Test-time attacks
    \item Jailbreaking techniques
    \item Attacks targeting LLM-based systems and agents rather than the underlying LLM itself
\end{itemize}

We first perform a title and abstract screening and remove papers clearly irrelevant to our review.  We next perform a full text review

\subsection{Data Extraction}
\textbf{General:}
\begin{itemize}
    \item \textbf{Year, publication source/venue}
    \item \textbf{Specific LLM models attacked}
    \item \textbf{Datasets used for each attack}
    \item \textbf{Computational resources used/required}: num GPUs, types of GPUS
\end{itemize}

\textbf{Attack model specifics recorded:}
\begin{itemize}
    \item \textbf{Subtlety consideration:} Is there any consideration for the trigger to avoid detection? (yes/no)
    \item \textbf{Clean label attack}: Is it a "clean label" attack, meaning the poisoned data must pass manual human inspection that the label is correct? (yes/no)
    \item \textbf{Efficiency Constraint}: Do the authors evaluate the "efficiency" of the poison? i.e. how much training data must be poisoned? (yes/no)
    \item \textbf{Amount of data poisoned} (Reported as percentage)
    \item \textbf{Task for poisoning}: (multiple choice)
    \begin{itemize}
        \item Pretraining
        \item Fine-tuning
        \item Multi-modal
        \item RLHF
        \item RAG
    \end{itemize}
    \item \textbf{Attacker-trained Model:} Did the attacker train the poisoned model themself? (yes) or provide the poisoned data for the model to be trained? (no)
    \item \textbf{Task / Application}: What was the model originally trained to do?
    \item \textbf{Evaluation of persistence through additional training / fine-tuning}(Yes/No)
    \item \textbf{Evaluation of persistence against defenses}: (Yes/No)
    \item \textbf{Defenses evaluated against}: (List of defenses used)
    \item \textbf{Evaluation of persistence across tasks}: (yes/no)
    \item \textbf{Evaluation of detectability}: Does the author evaluate if people or an algorithm can detect the poison placement on data or the poison behavior? (yes/no)
    \item \textbf{Evaluation attack success rate}: (Percentage of successful attacks)
    \item \textbf{Task success pre-poisoning}: The model's performance on the original task when trained without a poison (or before poisoning if poisoning fine-tuning).
    \item \textbf{Task success post-poisoning}: Performance on the original task after poisoning is present. 
\end{itemize}

\textbf{Poison Set Specifics}
\begin{itemize}
    \item \textbf{Paper-defined set name/type}: How does the paper refer to the set of points that they apply the poison to? Often called one-to-one or one-to-all to refer to only play to a target label.
    \item \textbf{Granularity of Change} (multiple choice)
    \begin{itemize}
        \item \textbf{Token Level - Character}
        \item \textbf{Token Level - Word}
        \item \textbf{Token Level - Subword} 
        \item \textbf{Concept Level}
        \item Other
    \end{itemize}
    \item \textbf{Change Types} (multiple choice)
    \begin{itemize}
        \item \textbf{Substitution}
        \item \textbf{Insertion}
        \item \textbf{Deletion}
    \end{itemize}
    \item Notes - (free text)
\end{itemize}

\textbf{Poison Behavior}
\begin{itemize}
    \item \textbf{Paper-defined behavior}: Often called "targeted" or "untargeted" refers to the desired change in the output of the model in the presence of a poison.
    \item \textbf{Single Output Label} (multiple choice)
    \begin{itemize}
        \item \textbf{Untargeted}: Any incorrect output label is valid
        \item \textbf{Targeted}: A specific, or set of specific, incorrect labels is a successful attack for each poisoned point. 
    \end{itemize}
    \item \textbf{Multi Output}
    \begin{itemize}
        \item \textbf{Level of Poison Behavior} (multiple choice)
        \begin{itemize}
            \item \textbf{Global}: Change the entire output of a specific input
            \item \textbf{Local}: Change only specific subsections of the output
        \end{itemize}
        \item \textbf{Type of change} (multiple choice)
        \begin{itemize}
            \item \textbf{Substitution}
            \item \textbf{Insertion}
            \item \textbf{Deletion}
        \end{itemize}
    \end{itemize}
    \item Notes - (free text)
\end{itemize}


\end{document}